# Phonon-Spectrum Narrowing Induced by Ultrafast Charge Fluctuation in an Organic Dimer Mott Insulator

K. Itoh[1], H. Itoh[1,2], S. Saito[3], I. Hosako[3], Y. Nakamura[4], H. Kishida[4], N. Yoneyama[5,2], T. Sasaki[5,2], S. Ishihara[1], and S. Iwai[*1,2]

[1]Department of Physics, Tohoku University, Sendai 980-8578, Japan

[2]JST, CREST, Sendai, 980-8578, Japan

[3]National Institute of Information and Communications Technology, Kobe 651-2492, Japan

[4]Department of Applied Physics, Nagoya University, Nagoya 464-8603, Japan

[5]Institute for Materials Research, Tohoku University, Sendai 980-8577, Japan

* Author to whom correspondence should be addressed

**Abstract** We have observed the characteristic temperature dependence of the intermolecular phonon spectrum in the organic dimer Mott insulator $\kappa$-(ET)$_2$Cu$_2$(CN)$_3$ exhibiting a dielectric anomaly at ~30 K. The anomalous spectral narrowing of the 55 cm$^{-1}$ phonon peak at ~30 K was analyzed in terms of motional narrowing within the framework of a stationary Gaussian process, i. e., the phonon frequency is modulated by the ultrafast charge fluctuation. The spectral narrowing occurs because the time constant of the correlation time $\tau_c$ and the amplitude $\Delta$ of the frequency modulation satisfy the relation $\tau_c < 1/\Delta$ at 30 K. At temperatures below 30 K, the motional narrowing is disturbed by the increasing of $\tau_c$, near the charge-glass or the short-range order at 6 K. On the other hand, for temperatures above 30 K, the motional narrowing is disturbed by the increase of $\Delta$ with increasing temperature.

**Introduction**

Electron ferroelectricity is a new feature of strongly correlated electron systems [1–5]. In some low-dimensional organic molecular salts and transition metal oxides such as (TMTTF)$_2$X (TMTTF; bis-tetramethyl-tetrathiafulvalene) [6], $\alpha$-(ET)$_2$I$_3$ (ET(BEDT-TTF); bis-ethylendithio-tetrathiafulvelen) [7, 8], $\kappa$-(ET)$_2$Cu[N(CN)$_2$]Cl [9], and LuFe$_2$O$_4$[10–13], inversion symmetry is broken by a strong Coulomb repulsion interaction. However, spatial symmetry breaking is sometimes limited in the microscopic region, resulting in the relaxor ferroelectric or polar nano-region (PNR) as demonstrated in organic dimer Mott insulators [14–18]. Large fluctuation and/or inhomogeneity of the short-range correlation of the charges are very important in such a correlated electron system exhibiting ferroelectricty or a dielectric anomaly (hereafter we refer to such compounds as correlated electron dielectrics). These fluctuating and inhomogeneous characteristics enable us to expect large and ultrafast responses in various electric and/or magnetic properties such as large magneto-electric effects, nonlinear conduction, and photoinduced phase transitions.

Very recently, collective excitation of electric dipoles with large fluctuation has been observed in the 1 THz (30 cm$^{-1}$) region [19] in the organic dimer Mott insulator $\kappa$-(ET)$_2$Cu$_2$(CN)$_3$ [20–23] as shown in Fig. 1(a), exhibiting a relaxor-like dielectric anomaly which has been attributable to the intradimer charge disproportionation (CD) or charge imbalance with large fluctuations [14–18]. Conventional experimental techniques such as Raman/infrared vibrational spectroscopy and NMR measurement, which are sensitive to the molecular charge, have not detected a clear peak splitting reflecting the CD. However, the anomalous spectral broadenings even at low temperature in those measurements [24, 25] suggest that the fluctuating and the imhomogenous short-range order or the glass state are formed in the vicinity of the FCO boundary in the theoretically considered phase diagram, as shown in Fig. 1(b) [16].

Recent studies on other compounds also indicate the importance of this unconventional CD [9, 15] and how such a state should be probed. The optical conductivity in the energy range above 100 cm$^{-1}$ indicates that the fluctuating charge/magnetic states in $\kappa$-(ET)$_2$Cu$_2$(CN)$_3$ are different from the ordinary antiferromagnetic Mott insulator [26]. Furthermore, our previous paper [19] demonstrated that measuring the ~30 cm$^{-1}$ band, which reflects the collective excitation of the fluctuating charge, is a powerful method for detecting such a state.

On the other hand, observation of the intermolecular low-frequency (<100 cm$^{-1}$) phonons is another effective approach for clarifying the nature of the fluctuating charge, because intermolecular charge motion interacts with the intermolecular phonons. However, the phonon dynamics in the correlated electron dielectrics remain unclear, although those of the conventional displacive-type ferroelectrics have been extensively studied; that is, symmetry breaking is induced by atomic displacements and infrared phonon modes such as soft modes, and a central mode shows characteristic behaviors reflecting a dielectric transition in the MHz and GHz frequency regions [27]. Different phonon dynamics are also expected to be shown at higher frequencies such as in the THz region in correlated electron dielectrics, because of the interaction between the phonon and the ultrafast charge fluctuation.

Here we use THz time-domain spectroscopy (TDS) to investigate the phonon dynamics of the organic dimer Mott insulator $\kappa$-(ET)$_2$Cu$_2$(CN)$_3$ exhibiting a relaxor-like dielectric anomaly. The 55 cm$^{-1}$ phonon spectrum observed for E//b polarization shows characteristic spectral narrowing at ~30 K. The change of the spectral shape of the 55

cm$^{-1}$ peak is attributable to the modulation of the phonon energy, which is driven by the ultrafast charge or dipole fluctuation. This is referred to as motional narrowing. At temperatures above and below 30 K, different mechanisms disturb the motional narrowing; these mechanisms are the increasing of the correlation time (< 30 K) and the increase of the amplitude of the fluctuating charges (> 30 K), respectively.

**Experimental**

Single crystals of $\kappa$-(ET)$_2$Cu$_2$(CN)$_3$ (average size: 1 × 1 × 0.5 mm$^3$) were grown by an electro-chemical method [14]. Raman spectra were measured by a Renishaw inVia Raman spectrometer equipped with a NExT filter unit. The excitation wavelength is 632.8 nm. The THz response was measured by THz TDS in the energy range of 10–100 cm$^{-1}$ in the transmission configuration, using a Tochigi Nikon Rayfact spectrometer. Generation of the THz pulse was performed by using a 100 fs fiber laser to excite a photoconducting switch under a bias. The THz pulse transmitted through the sample was also detected with a photoconducting switch. The obtained time-domain profile is extrapolated to zero after the modulating signal has completely decayed at $t_{ext}$= 80 ps. The frequency resolution of the transmittance $t(\omega)$ spectrum is directly determined by the time span of the time-domain measurement (~0.4 cm$^{-1}$ for $t_{ext}$= 80 ps). However, the spectral shape in the optical conductivity $\sigma(\omega)$ shows the sharp peaks, which are narrower than the bandwidth of $t(\omega)$, because $t(\omega) \propto \exp(-k(\omega))$, whereas the shape of the optical conductivity spectrum is described by the relation $\sigma(\omega) \propto k(\omega)$, where $k$ represents extinction coefficient. As a result, the ~ 0.1 cm$^{-1}$ peak can be detected in the $\sigma(\omega)$ spectra.

**Results**

Figures 2(a) and 2(b) show the Raman spectrum of $\kappa$-(ET)$_2$Cu$_2$(CN)$_3$ at 10 K, covering the spectral range of 27–230 cm$^{-1}$ (Fig. 2(a)) and 27–115 cm$^{-1}$ (Fig. 2(b)), respectively. The polarizations of the excitation and detection are both parallel to the *b* axis (bb polarization). Figures 2(c) and 2(d) show the $\sigma(\omega)$ spectra for polarizations parallel to the *b* axis (E//b) and the *c* axis (E//c), respectively [19]. Several strong peaks were observed below 100 cm$^{-1}$ in both the Raman and $\sigma(\omega)$ spectra. It is noteworthy that the intensity of the Raman signals below 100 cm$^{-1}$ is distinctively enhanced. Accordingly, the intramolecular modes (> 100 cm$^{-1}$) and intermolecular modes (< 100 cm$^{-1}$) can be distinguished easily, considering that the frequencies of the lowest intramolecular modes of the isolated ET molecule are evaluated to be 258 cm$^{-1}$ (for IR 2$_u$) and 161 cm$^{-1}$ (for Raman A$_g$) [28], respectively. Therefore, the spectral features below 100 cm$^{-1}$ are essentially attributable to the intermolecular modes, although such low-frequency peaks include an intramolecular-mode component [29]. Since the peak energies at 37, 45, 53, and 70 cm$^{-1}$ in the Raman spectrum are close to those for the $\sigma(\omega)$ spectra (39, 41, 55, and 69 cm$^{-1}$) at 10 K, respectively, the oscillating modes below 100 cm$^{-1}$ observed in the $\sigma(\omega)$ spectra are attributable to the intermolecular modes which are similar to those detected in the Raman spectrum. That is because the corresponding infrared(IR) active and Raman active modes, i.e., an in-phase motion (Raman) and an out of phase motion (IR) of the molecules in a symmetric configuration, give approximately equal frequencies. We also noticed that the broad electronic background for E//c (~70 $\Omega^{-1}$cm$^{-1}$ at 80 cm$^{-1}$) in Fig. 2(d) is much larger than that for E//b (~6.5 $\Omega^{-1}$cm$^{-1}$ at 80 cm$^{-1}$) in Fig. 2(c). Furthermore, the spectrum for E//c exhibits a dip and/or a marked asymmetry that is attributable to a Fano interference between the sharp phonon peaks and the electronic background. Such spectral asymmetry of the phonon peaks reflecting the Fano effect is also detected for E//b, although it is smaller than that for E//c. These anisotropic features in the optical conductivity spectrum are consistent with the theoretical analysis; that is, the dielectric response reflecting the intradimer CD can occur in the E//c direction [30]. Therefore, the possible reason for the spectral anisotropy is the intradimer dipoles. In particular, the broad peak at 30 cm$^{-1}$ with the Fano-interference dip shows a characteristic temperature dependence [19], indicating that the 30 cm$^{-1}$ band is closely related to a previously observed dielectric anomaly [14]. Based on the temperature dependence of this 30 cm$^{-1}$ band and theoretical considerations [16, 30], the 30 cm$^{-1}$ band is attributable to the collective excitation of the intradimer electric dipole. It is noteworthy that this collective excitation is detected only for E//c, which is a result also supported by the theory.

In contrast to the E//c spectrum with a large background and a strong Fano effect, clear phonon peaks with a small background and an asymmetry are observed at 23, 39, 41, 55, 69, 84, 87, and 92 cm$^{-1}$ for E//b, as shown in Fig. 2(c). The temperature dependence of the normalized peak energies $\omega_0/\omega_{0(5\ K)}$, the spectral bandwidth (the full width at half maximum (FWHM)), and the peak intensity are respectively shown in Figs. 3(a), 3(b), and 3(c). They were obtained numerically after the backgrounds (as shown by the dashed line in Fig. 3(e)) were subtracted. The temperature dependences of the background conductivity at 40 cm$^{-1}$, 55 cm$^{-1}$, and 69 cm$^{-1}$, which are essentially consistent with the earler result [26], are shown in Fig. 3(d). Small (<2 %) monotonic increases (38, 41, 55, and 69 cm$^{-1}$) of $\omega_0$ with decreasing temperature were observed (along with a small monotonic decrease of $\omega_0$ at 23 cm$^{-1}$), reflecting the anisotropic thermal expansion and shrinkage of the lattice as previously reported [32]. We cannot find any dielectric responses indicating relaxation modes associated with the ferroelectric soft mode and/or the central mode in this energy region. That is in contrast with the fact that the dielectric measurement in the region from several hundreds of Hz to MHz has detected the relaxation mode [14], reflecting the macroscopic domain motion. On the other hand, the spectral bandwidth of the ~55 cm$^{-1}$ peak for E//b shows anomalous narrowing at ~30 K, as shown in Fig. 3(b), in contrast with the monotonic change of the peak energies; that is, with increasing temperature, the bandwidth decreases for temperatures below 30 K and increases for temperatures above 30 K. It is noteworthy that only the 55 cm$^{-1}$ peak exhibits such anomalous narrowing. Spectra of the ~55 cm$^{-1}$ phonon peak observed for various temperatures are shown in Fig. 3(f).

Here, we introduce a normalized spectral line-shape function $I_{norm}(\omega)$ which was normalized by both bandwidth and peak intensity after subtracting the background. The normalized spectral line-shape function in Fig. 4 was obtained by the following procedure: i) The peak intensity was normalized after subtracting the background (the manner in which we subtract the background is shown by Fig. 3(e)), ii) the peak energy was shifted to 0 cm$^{-1}$, and iii) the bandwidth (FWHM) was normalized. The peak energy, the bandwidth, and the peak intensity have been already presented in Figs. 3(a), 3(b), and 3(c). Figure 4 shows the $I_{norm}(\omega)$ for (a) 30 K and (b) 5 and 80 K. We note that the spectrum at 30 K (Fig. 4(a)) has large spectral tails for both sides, although these tails are apparently smaller at 5 K and 80 K (Fig. 4(b)), suggesting a change in the spectrum from a Gaussian shape to a Lorentzian shape as shown in Figs. 4(c)–(e) (which were calculated by Eq. (5), as described later). The spectral asymmetry in Fig. 4(a) is attributable to Fano effects, as shown above.

**Discussion**

Based on the fact that the charge fluctuation, which might be related to the intradimer dipole, becomes active below ~30 K [14, 19], the spectral narrowing and the change from the ~55 cm$^{-1}$ phonon spectrum with small tails (5 and 80 K) to that with large tails (30 K) are attributable to the modulation of the phonon energy, which is driven by the ultrafast charge fluctuation. This is due to the motional narrowing [33]. Here, we report our analysis of the spectral shape and width in the framework of a stationary Gaussian process [33].

Modulating the phonon energy $\omega_0$ by the random fluctuation $\omega_f(t)$, the time evolution of the angular frequency of the phonon is written as $\omega(t) = \omega_0 + \omega_f(t)$. In the framework of a Gaussian process, the correlation function of the phonon displacement $x$ is described by

$$\langle x^*(0) x(t) \rangle = e^{i\omega_0 t} \phi(t) \tag{1}$$

$$\phi(t) = \left\langle \exp\left( i \int_0^t \omega_f(t') dt' \right) \right\rangle, \tag{2}$$

for a random initial value of $x$. The equation $\langle \omega_f(t) \rangle = 0$ is satisfied for the stationary frequency modulation process, and the correlation function of $\omega_f$ is described by using $\psi(t)$, representing the time evolution of the correlation function,

$$\langle \omega_f^*(t_0) \omega_f(t_0 + t) \rangle = \langle \omega_f^2 \rangle \psi(t). \tag{3}$$

If we assume $\psi(t) = \exp(-t/\tau_c)$, where $\tau_c$ is the time constant of the random fluctuation, $\phi$ can be calculated as follows,

$$\phi(t) = \exp\left[ -\alpha^2 \left( \frac{t}{\tau_c} - 1 + e^{-t/\tau_c} \right) \right]. \tag{4}$$

$\alpha \equiv \tau_c \Delta$ is introduced as the line-shape parameter, where $\Delta^2 \equiv \langle \omega_f^2 \rangle$ is the amplitude of the fluctuation. Here, the phonon spectrum $I(\omega)$ is given by the Fourier transform of the correlation function $\phi(t)$ in the framework of the Wiener–Khinchin theorem:

$$I(\omega) = \frac{1}{2\pi} \int_{-\infty}^{\infty} \phi(t) e^{-i\omega t} dt. \qquad (5)$$

We fitted the observed spectra for various temperatures by using $I(\omega)$ (Eq. (5)), as shown in Fig. 5(a).

As shown by the red curves in Fig. 5(a), the observed spectra can be roughly reproduced by the formulation above. This indicates that the narrowing of the 55 cm$^{-1}$ phonon spectrum involves a change of the line-shape from one with a small tail (5 and 80 K) to a shape with a large tail (30 K). Figures 5(b), 5(c), and 5(d) show the temperature dependences of the line-shape parameter $\alpha$, the time constant $\tau_c$, and the amplitude of the fluctuation $\Delta$, respectively. Figure 5(a) shows that the spectral narrowing can be characterized by a marked decrease of $\alpha$ at 30 K. For 5–30 K, as shown in Fig. 5(b), the decrease of $\alpha$ with increasing temperature is attributable to the reduction of $\tau_c$ to 0.4 ps at 30 K. On the other hand, the increase of $\alpha$ for temperatures above 30 K mainly comes from the increase of $\Delta$ at high temperature, as indicated by Fig. 5(d). As shown above, the 55 cm$^{-1}$ phonon peak shows a small asymmetry induced by the Fano effect. However, the obtained parameters are not seriously affected by the Fano effect, because the asymmetry is very small.

The possible reasons for the anomalous changes in $\tau_c$ and $\Delta$ are considered to be as follows: the relaxor-like behavior of the dielectric constant below 30 K with large frequency dispersion [14] has been attributed to the fluctuating microscopic CD or electric dipole at ~30 K and formation of the glass state or short-range dipole order at temperatures below 6 K. Taking this into consideration, the anomalous increase of $\tau_c$ from 30 K to 6 K reflects the fact that the fluctuation of the microscopic CD becomes slower near the formation of the glass state or the short-range order at 6 K. On the other hand, for temperatures above 30 K, $\Delta$ increases with temperature. In other words, $\Delta$ decreases toward 30 K as temperature decreases from 80 K, which is attributable to the softening of the charge fluctuation. Such softening is suppressed at low temperature (< 30 K), reflecting the fact that the charge correlation does not grow to form long-range ferroelectric charge order, resulting in microscopic or short-range polar domains. According to the phase diagram obtained by the theory [16], the possible mechanism for this suppression of the ferroelectric order below 30 K might be related to the competing charge and magnetic correlations [34–36].

The decrease of the bandwidth of the 55 cm$^{-1}$ peak toward 30 K at temperatures above 30 K also might include the effect that is caused by the suppression of the phonon-phonon scattering at low temperature. In fact, the widths of other phonon peaks (such as the 69 cm$^{-1}$ peak shown in Fig. 4) show the monotonic decrease which is attributable to the suppression of the phonon–phonon scattering. However, the characteristic temperature dependence of $\Delta$ shown in Fig. 5(d) suggests that other origins, such as the softening of the charge fluctuation, are dominant.

Thus, the narrowing of the 55 cm$^{-1}$ phonon spectrum at 30 K is attributable to the scattering between the phonon and the ultrafast charge modulation. For $T < 30$ K, the spectral shape is mainly determined by $\tau_c$, whereas it is governed by $\Delta$ for $T > 30$ K. The "V"-like shape observed in the temperature dependence of the spectral width (Fig.

3 (b)) and in that of the line-shape parameter $\alpha$ (Fig. 5(b)) are attributable to the increase of $\tau_c$ near 6 K and the softening of $\Delta$ with decreasing temperature, although the effect of phonon–phonon scattering might be included. Such anomalous temperature dependences of $\tau_c$ and $\Delta$ come from the fluctuating and inhomogeneous character of the charge: i) the condition $\tau_c < 1/\Delta$ for the spectral narrowing is satisfied because of the ultrafast charge motion, ii) $\tau_c$ increases below 30 K, but does not show the "critical" behavior, because any long-range order is not formed below 6 K, and iii) $\Delta$ shows softening with decreasing temperature but is suppressed below ~30 K, because of the fluctuations. As mentioned before, the small $\alpha$ value ($< 0.2$) at ~30 K corresponds to a short $\tau_c$ (0.4 ps). It is noteworthy that the value of $\tau_c$ at ~30 K (0.4 ps) is roughly equal to the time scale of the collective motion, which has been evaluated from the collective excitation energy (1/(1 THz) ~ 0.9 ps)) [19].

It is important to discuss about the reason that the motional narrowing is observed only for the 55 cm$^{-1}$ mode. We can definitely state that such low-frequency vibration corresponds to the intermolecular (optical phonon) mode. The intradimer stretching or libration modes modulating the intermolecular transfer integral are possible candidates for the origin of the 55 cm$^{-1}$ mode, although it is difficult to show the reliable assignment of low-frequency phonons ($< 100$ cm$^{-1}$) in a two-dimensional organic system containing many atoms in the unit cell [29]. If so, such phonon modes should selectively interact with the intermolecular (intradimer) charge fluctuation.

In the previous papers, neither definite long-range charge nor magnetic order were found to form below 6 K in these compounds. However, several indications have been observed, showing the changes of the electronic/magnetic properties at 6 K in the thermodynamic nature [37, 38], lattice properties [32], dielectric properties [14], and magnetic properties [21, 36]. Furthermore, our recent results on the ∼30 cm$^{-1}$ response (for the E//c axis) also show the anomaly at 6 K [19]. These results shows that the fluctuating and inhomogeneous short-range order or the charge-glass state are formed below 6 K. Considering that, our results are consistent with the scenario, i. e., the spectral shape of the phonons is affected by the ultrafast charge fluctuation.

In summary, we observed characteristic spectral narrowing of the intermolecular 55 cm$^{-1}$ mode phonon (E//b) in $\kappa$-(ET)$_2$Cu$_2$(CN)$_3$ by using THz TDS. Assuming a stationary Gaussian process for the modulation of the phonon energy, the spectral narrowing at 30 K and the changing of the spectral shape with small tails (80 K) to the large-tailed shape (30 K), and then back to the small-tailed one (10 K) are attributable to the ultrafast modulation of the phonon energy driven by the intradimer dipole fluctuation through the motional narrowing effect. At temperatures above and below 30 K, different mechanisms disturb the motional narrowing; these mechanisms are the increasing of the time constant of the correlation time (< 30 K) and the softening of the modulating amplitude (> 30 K), respectively.

**Figure caption:**

**FIG. 1** Schematic illustration of **(a)** the crystal structure of $\kappa$-(ET)$_2$Cu$_2$(CN)$_3$ and **(b)** the theoretically predicted phase diagram of a dimer Mott insulator $\kappa$-ET salt. The molecular arrangement and charge distribution of the dimer Mott phase and ferroelectric CO phase are also shown. The red arrow (*P*) shows the direction of the macroscopic polarization *P* in the ferroelectric CO phase.

**FIG. 2** Raman spectrum of $\kappa$-(ET)$_2$Cu$_2$(CN)$_3$ at 10 K, covering the spectral range of 27–230 cm$^{-1}$ (Fig. 2(a)) and 27–115 cm$^{-1}$ (Fig. 2(b)), are respectively shown. The polarization of the excitation and detection are both parallel to the *b* axis (bb). The optical conductivity ($\sigma(\omega)$) spectra at various temperatures (10–70 K) for E//b **(c)** and E//c **(d)**.

**FIG. 3** Temperature dependence of **(a)** the normalized phonon peak energies $\omega_0/\omega_{0(5\,K)}$, **(b)** the spectral bandwidth (specifically, the full width at half maximum (FWHM)), **(c)** the peak intensities, and **(d)** the backgrounds at 40, 55, and 69 cm$^{-1}$ for E//b. **(e)** The manner by which we subtract the background is shown by the dashed line in the $\sigma$ *s*pectrum(30 K, E//b). **(f)** The spectrum of the ~55 cm$^{-1}$ phonon peak observed for various temperatures.

**FIG. 4** Normalized spectral line-shape function $I_{\text{norm}}(\omega)$ that is normalized by the FWHM and the peak intensity after subtracting the background for **(a)** 30 K and **(b)** 5 and 80 K. The spectra shown in **(c), (d),** and **(e)** were calculated with Eq. 5 for **(a)** $\alpha$=0.25, **(b)** $\alpha$=1, and **(c)** $\alpha$=$\infty$, respectively.

**FIG. 5 (a)** Observed spectra for various temperatures are shown together with the calculated $I(\omega)$ (given by Eq. (5) in the text). **(b), (c), and (d):** Temperature dependences of **(b)** the line-shape parameter $\alpha$, **(c)** the time constant $\tau_c$, and **(d)** the amplitude of the fluctuation $\Delta$.

(a)

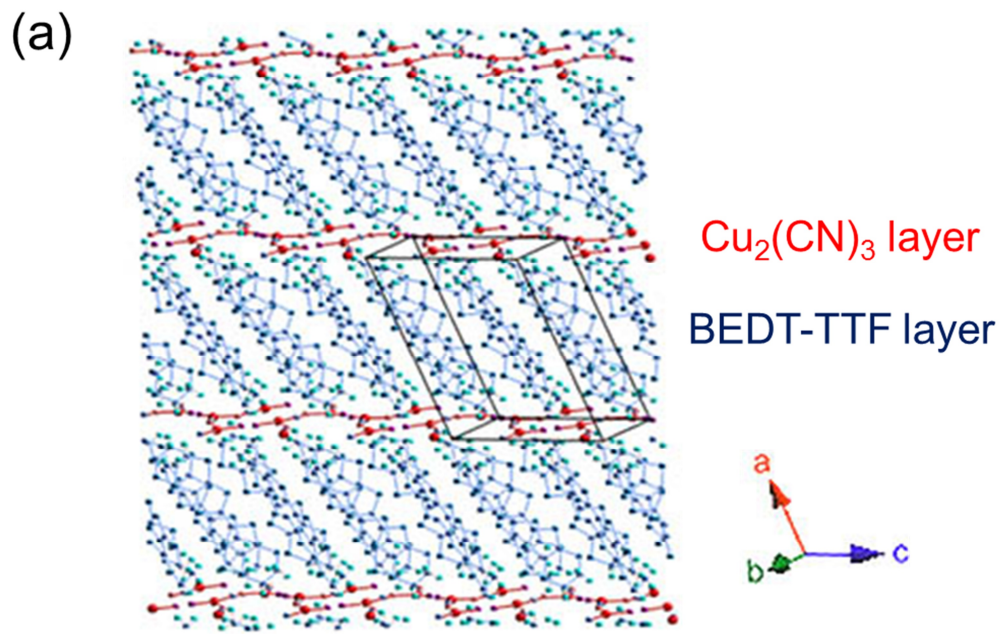

(b)

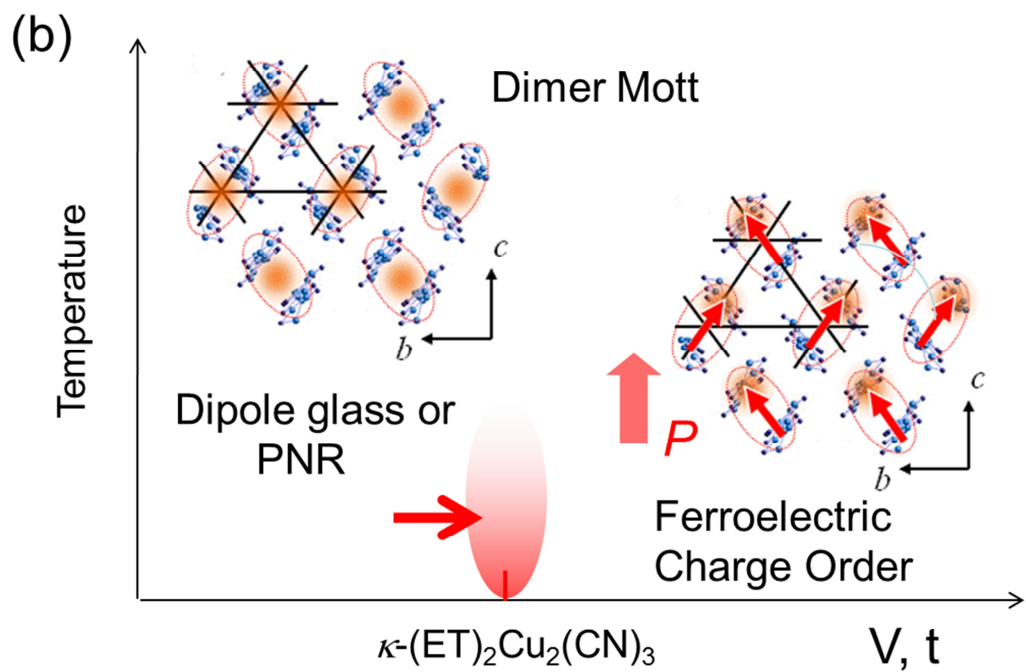

Fig. 1 Itoh et al

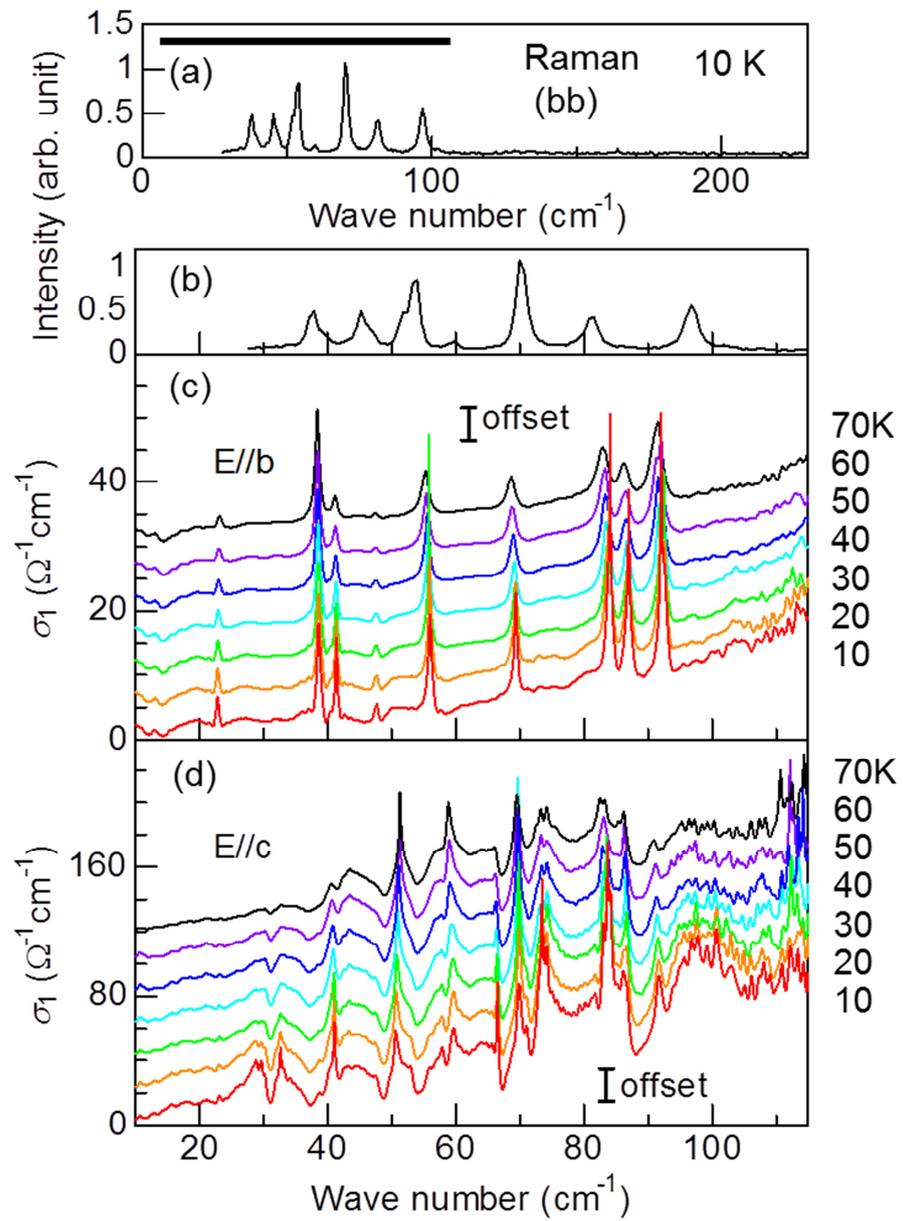

Fig. 2 Itoh et al

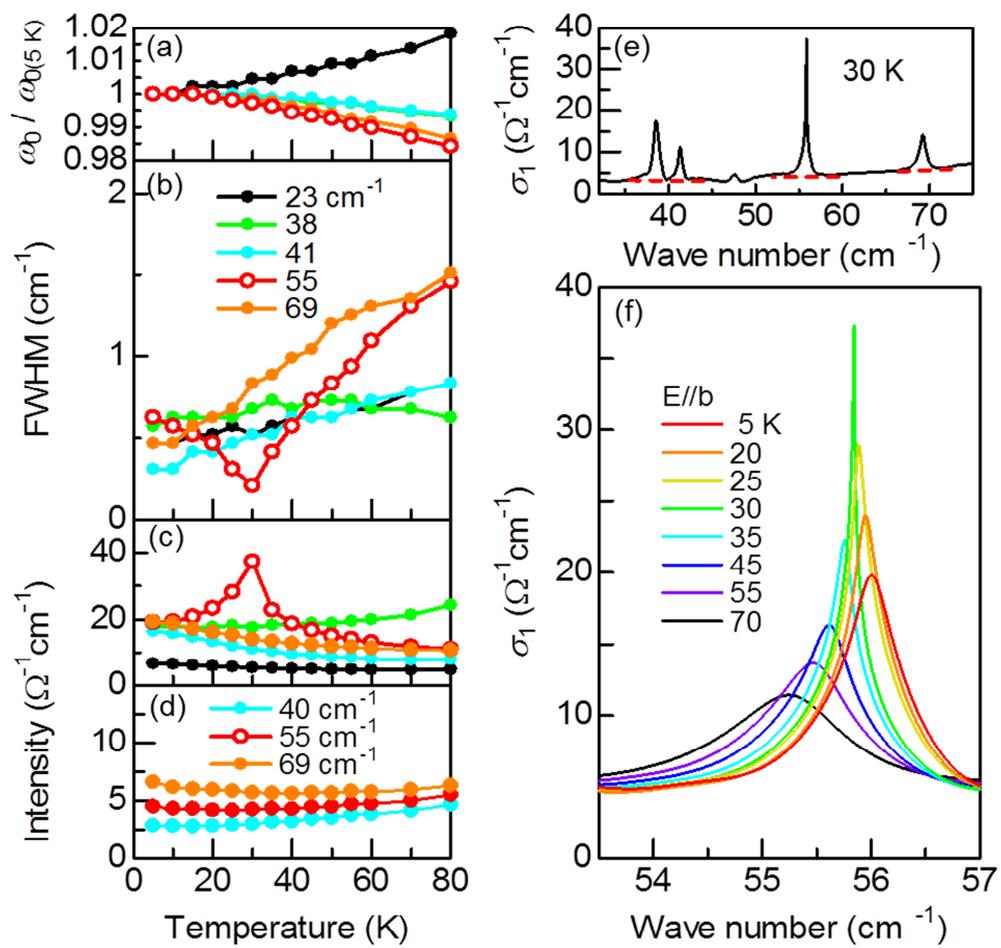

Fig. 3  Itoh et al

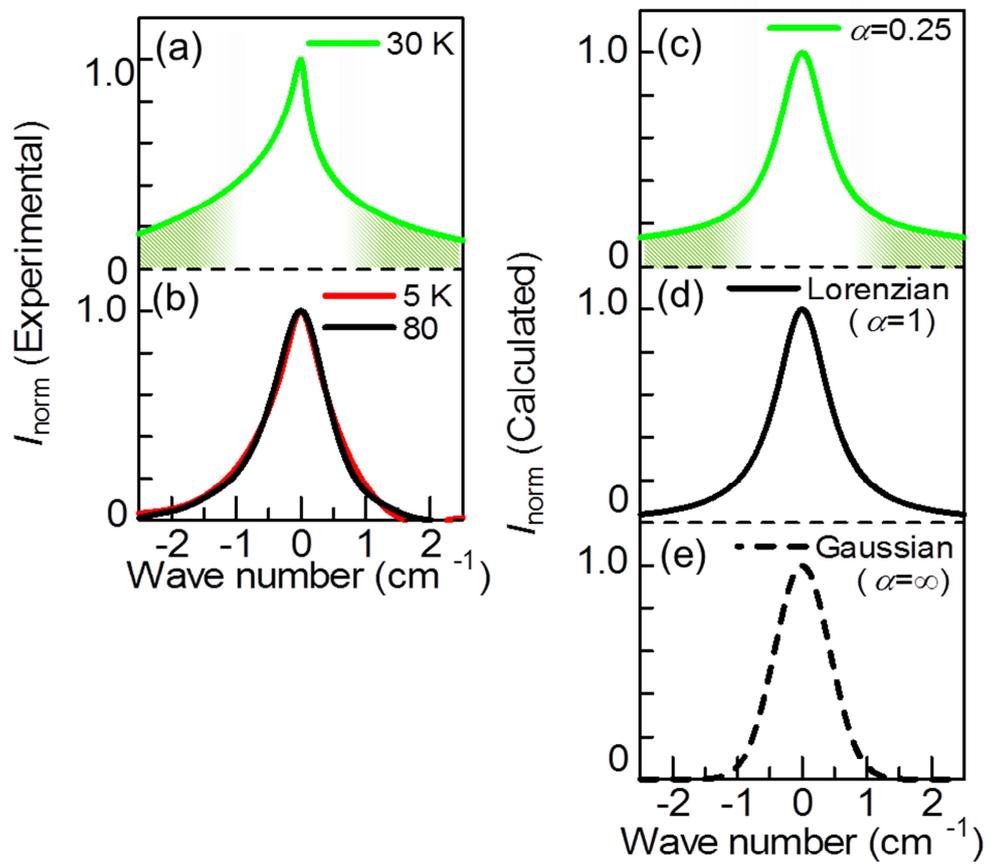

Fig. 4 Itoh et al

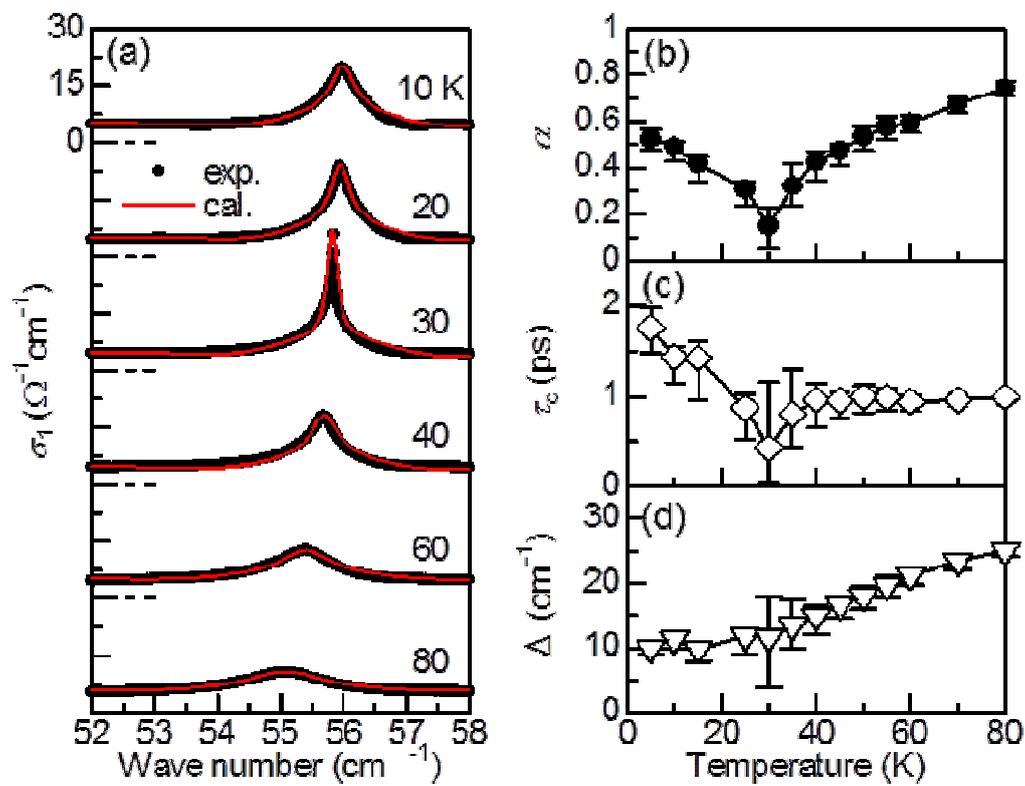

Fig. 5 Itoh et al